\documentstyle[twocolumn,aps]{revtex} 
 
\draft 
\begin{document} 
 
%%%%%%%%%%%%%%%%%%%%%%%%%%%%%%%%%%%%%%%%%%%%%%%%%%%%%%%%%%%%%%%%%%%%%%%%%%%%%%% 
 
\title{Low Field Anomaly in the Specific Heat of $s$-wave
Superconductors due to the Expansion of the Vortex Cores}

\author{J.E.~Sonier$^1$, M.F.~Hundley$^1$, J.D.~Thompson$^1$ and J.W.~Brill$^2$}
\address{$^1$Los Alamos National Laboratory,
Los Alamos, New Mexico 87545}
\address{$^2$Department of Physics and Astronomy, University of Kentucky,
Lexington, Kentucky 40506-0055}

\date{May 8, 1999} 
\date{ \rule{2.5in}{0pt} } 
 
\maketitle 
%%%%%%%%%%%%%%%%%%%%%%%%%%%%%%%%%%%%%%%%%%%%%%%%%%%%%%%%%%%%%%%%%%%%%%%%%%%%%%% 
\begin{abstract} \noindent 
The magnetic field dependence of the 
electronic specific heat $C(H)$ in the $s$-wave superconductor NbSe$_2$
shows curvature at low fields, resembling the near $\sqrt{H}$ term 
in $C(H)$ which
has been reported in high-$T_c$ superconductors and
attributed to a $d_{x^2-y^2}$-wave pairing state. In NbSe$_2$
we find that the low-field behaviour in $C(H)$ is described
quantitatively by the expansion of vortex cores 
and the field dependence of the magnetic induction $B$ above $H_{c1}$.
The associated change in the density of quasiparticle states localized
in the vortex cores provides a simple explanation for the
``low-field anomaly'' in $C(H)$ observed in $s$-wave superconductors.

\end{abstract} 
\pacs{ 74.25.Bt, 74.25.Jb, 74.70.Ad, 74.72.Bk} 
%%%%%%%%%%%%%%%%%%%%%%%%%%%%%%%%%%%%%%%%%%%%%%%%%%%%%%%%%%%%%%%%%%%%%%%%%%%%%%% 
%%\newpage 

In the vortex state of a type-II superconductor, an applied magnetic
field penetrates the bulk in the form of quantized flux lines.
The predictions of this state by Abrikosov \cite{Abrikosov:57}
and of bound quasiparticle (QP) states in the normal vortex cores
of a conventional superconductor by 
Caroli {\it et al.} \cite{Caroli:64} remain two great
achievements in superconductivity theory. Although flux decoration
experiments readily confirmed the existence of the vortex state, 
the electronic structure of the cores predicted by Caroli {\it et al.} 
was not established until the advent of scanning tunneling microscopy (STM). 
In particular, Hess {\it et al.} \cite{Hess:89} observed tunneling spectra 
in the vortex cores of NbSe$_2$ consistent with localized QP
states---thus seemingly completing the picture of the vortex state.

Interest in the vortex state was renewed with the discovery of the
high-$T_c$ superconductors (HTSCs). The general consensus is that the
charge carriers in the HTSCs form pairs whose wavefunction (or order
parameter) has a dominant $d_{x^2-y^2}$-wave symmetry, rather than 
the $s$-wave symmmetry characteristic of low-$T_c$ conventional 
superconductors. Much theoretical work has focussed on incorporating 
$d_{x^2-y^2}$-wave symmetry into a model for the vortex state of the HTSCs. 
Despite these efforts, experiments have yet to unambiguously confirm these 
predictions. 

Measurements of the electronic specific heat $C(T,H)$ are one way
of probing the QP excitation spectrum in the vortex state.
In an $s$-wave superconductor, where there is an isotropic energy gap at
the Fermi surface, there is a contribution to $C(H)$ which
is proportional to the QP density
of states (DOS) localized in the vortex cores. Since the density of
vortices increases linearly as a function of magnetic field, this term
is expected to be proportional to $H$ \cite{Fetter:69}. On the other
hand, in the vortex state of a $d_{x^2-y^2}$-wave superconductor,
Volovik \cite{Volovik:93} predicted that the DOS varies as $\sqrt{H}$.
This weaker field dependence of the DOS 
is mainly due to delocalized QPs 
which leak outside of the vortex cores along the nodal
directions of the order parameter. 
Within this model, bound core states are at most a minor correction
to the total DOS. Experiments performed on the HTSCs, 
YBa$_2$Cu$_3$O$_{7-\delta}$ (YBCO) 
\cite{Moler:94,Fisher:95} and La$_{2-x}$Sr$_{x}$CuO$_4$ (LSCO)
\cite{Khlopkin:97,Chen:98}, showed that there is a contribution to
$C(H)$ which is approximately proportional to $\sqrt{H}$---consistent with
Volovik's prediction. Although this was one of the key early experiments
providing evidence for $d_{x^2-y^2}$-wave symmetry in the HTSCs,
the interpretation of such measurements has been plagued by puzzling
reports of similar curvature in $C(H)$ at low magnetic fields 
in $s$-wave superconductors, such as NbSe$_2$ (Ref.~\cite{Sanchez:95}), 
V$_3$Si (Ref.~\cite{Ramirez:96}) and CeRu$_2$ (Ref.~\cite{Hedo:98})---and 
in other unconventional superconductors, like the heavy fermion superconductor 
UPt$_3$ (Ref.~\cite{Ramirez:95}),
the organic superconductor (BEDT-TFF)$_2$Cu[N(CN)$_2$]Br 
(Ref.~\cite{Nakazawa:97}) and the borocarbide superconductor 
LuNi$_2$B$_2$C (Ref.~\cite{Nohara:97}).
Ramirez \cite{Ramirez:96} suggested that this behavior
at low fields might be a general feature
of all superconductors in the vortex state, independent of the order
parameter symmetry, but somehow
related to the strength of the vortex-vortex interactions. 
Clearly the $\sqrt{H}$ dependence of the specific heat 
in the HTSCs cannot be attributed
to nodes in the energy gap function without a satisfactory explanation
for similar behavior in fully gapped superconductors.

A long-standing general belief is that the
superconducting coherence length $\xi$ is independent of $H$.
However, muon spin rotation ($\mu$SR) \cite{Sonier:97a}
and STM \cite{Hartmann:93} measurements have
shown that the radius $r_0$ of the vortex cores in NbSe$_2$ expand at
low fields. In the $\mu$SR experiment, the field dependence of $r_0$ 
could be fit to a function dependent only on the intervortex spacing. 
These $\mu$SR and STM results have been met with some skepticism,
however, since the change in $r_0$ implies that the superconducting
coherence length $\xi$ also varies with field. An obvious 
question raised by these observations is: 
Could the low-field anomaly in the specific heat be due to the expansion
of the vortex cores?

To address this question we have carried out detailed measurements of the
specific heat $C(T,H)$ in a 40~mg crystal of NbSe$_2$ from the same batch
used in the $\mu$SR experiment of Ref.~\cite{Sonier:97a}. The crystal
had a superconducting transition temperature
$T_c \! = \! 7.0(1)$~K, as determined previously \cite{Sonier:97a}
by resistivity and susceptibility measurements, and here from
measurements of the specific heat in zero field (see Fig.~1).
From magnetization and specific heat measurements, the upper and lower
critical fields for this crystal were found to be $H_{c2} \! = \! 2.90(3)$~T and
$H_{c1} \! = \! 0.025(2)$~T at $T \! = \! 2.3$~K, and
$H_{c2} \! = \! 1.75(2)$~T and $H_{c1} \! = \! 0.015(1)$~T at $T \! = \! 4.2$~K.
The specific heat was measured using a thermal relaxation calorimeter
\cite{Bachmann:72} with the magnetic field applied normal to the
NbSe$_2$ layer direction. The data have been corrected for
the small field-dependent background and addenda contributions.  

Figure~2 shows the field dependence of $C(T,H)/T$ in NbSe$_2$
at $T \! = \! 2.3$~K and below $H_{c2}$. 
The downward curvature at low fields is similar 
to that reported in Ref.~\cite{Sanchez:95}. Measurements were taken
four different ways: ({\it i}) The single crystal was zero-field 
cooled (ZFC) to $T \! = \! 2.3$~K, and measurements taken for 
increasing field up to $H \! = \! 2.5$~T,
({\it ii}) then taken for decreasing field down to
$H \! = \! 0$~T. ({\it iii}) Starting with the crystal at $T \! = \! 2.3$~K and 
$H \! = \! 4$~T ({\it i.e.} in the normal state), the field was ramped down to 
$1$~T and measurements taken for decreasing field down to $H \! = \! 0$~T.
({\it iv}) Measurements were also made after field-cooling (FC) the 
crystal to $T \! = \! 2.3$~K, in different magnetic fields. 
No significant hysteresis is found in the specific heat
measurements---consistent with the near reversibility of the
magnetization (Fig.~2, inset). 
    
In a conventional $s$-wave superconductor, the specific heat
in the vortex state ($H \! > \! H_{c1}$) is greater 
than in the Meissner state ($H \! < \! H_{c1}$), 
due to a contribution from the localized QPs in the vortex cores.
A precise calculation, from the Bogoliubov equations
assuming noninteracting vortices, gives the density of states $N(E)$
per unit volume associated with the bound excitations as \cite{Fetter:69}
\begin{equation}
N(E) = N(0) \pi \xi_0 \left[ \frac{\pi I}{f^\prime} \right]
\frac{B(H)}{\Phi_0} \, ,
\label{eq:DOS}
\end{equation}
where $\xi_0$ is the coherence length, $N(0)$ is the density of normal 
electron states at the Fermi surface, $f^\prime$ is the slope of the
order parameter at the vortex axis $r \! = \! 0$, $I$ is a numerical constant,
$\Phi_0$ is the flux quantum and $B(H)$ is the magnetic induction. 
The factor $B(H)/\Phi_0$ is the density of vortices. 
The coherence length in Eq.~(\ref{eq:DOS}) is related
to the value of the energy gap $\Delta_0$ far outside the vortex cores
{\it i.e.}, $\xi_0 \! = \! \hbar v_F/\pi \Delta_0$.
In Ginzburg-Landau (GL) theory, $f \! \approx \! r/ \xi$ at $r \! = \! 0$, so that 
$f^\prime \! \approx \! 1/\xi$ \cite{Caroli:64,Tinkham:96}.
The GL coherence length $\xi$ is the characteristic length scale
for spatial variations in the order parameter, and unlike
$\xi_0$, depends on temperature and is proportional
to the field-dependent vortex-core radius.
Using Eq.~(\ref{eq:DOS}), the contribution of the vortex cores
to the specific heat when $k_B T \gg \Delta_0^2/E_F$ is  
\begin{equation}
C_{\rm cores}(T,H) = \frac{2}{3} \pi^2 N(E) k_B^2 T
= \pi^2 \xi_0 \gamma_n T \left[ \frac{B(H)}{\Phi_0}
\right] \xi I \, , 
\label{eq:Fetter}
\end{equation}
where $\gamma_n T \! = \! (2/3) \pi^2 N(0) k_B^2 T$ 
is the normal-state electronic specific heat and
$E_F$ is the Fermi energy. Thus, the specific heat in the vortex 
state may be written as
\begin{equation}
C(T,H) = C(T,0) + \pi^2 \xi_0 \gamma_n T \left[ \frac{B(H)}{\Phi_0}
\right] \xi I \, , 
\label{eq:heat}
\end{equation}
where $C(T,0)$ is the specific heat at zero field in the Meissner
state---with values from Fig.~1 given in Table~\ref{fits}.

Maki \cite{Maki:65} was the first to consider the effect of
vortex-vortex interactions on the specific heat---deriving
the following relation from London theory, just above $H_{c1}$ 
in the high-$\kappa$ limit 
\begin{eqnarray}
C(T,H) & = & C(T,0) - \frac{T B}{4 \pi} \frac{d^2H_{c1}}{dT^2}
 + \frac{TB^{3/2} \lambda}{H-H_{c1}} 
\left( \frac{dH_{c1}}{dT} \right)^2 \nonumber \\ 
& \times & \left( \frac{\sqrt{3}}{8 \pi^2 \Phi_0} \right)^{1/2} 
+ \hbox{\rm terms in} \, 
\frac{1}{\lambda} \frac{d \lambda}{dT} \, ,
\label{eq:Maki}
\end{eqnarray}
where $\lambda$ is the magnetic penetration depth. 
Strictly speaking, $B$ and $\lambda$ are functions of both $T$
and $H$. The second term in Eq.~({\ref{eq:Maki}) has the same 
form as Eq.~(\ref{eq:Fetter}), whereas the third term describes the
perturbation due to vortex-vortex interactions. 
In Ref.~\cite{Ramirez:96} it was suggested that this latter term 
may explain the downward curvature in $C(H)$ observed at low fields
in V$_3$Si. However, Eq.~(\ref{eq:Maki}) has never been
quantitatively verified. Furthermore, it was noted in
earlier measurements of $C(T,H)$ in V$_3$Si (Ref.~\cite{Brock:69}),
that Eq.~(\ref{eq:Maki}) does not yield a term $\sim HT^3$
which is observed experimentally.

Recently, Ichioka {\it et al.} \cite{Ichioka:99} used the quasiclassical Eilenberger
equations to calculate the DOS in an $s$-wave superconductor with interacting
vortices. At $T \! = \! 0$~K the DOS is found to vary as $H^{0.67}$ due to a change in
the slope of the order parameter in the vortex cores.
Fitting the increasing field ZFC data for $C(T,H)/T$ 
in the inset of Figs.~3 and 4 (the fit is not shown) to a function
of the form $C(T,0)/T \! + \! \beta H^n$ using the values of
$C(T,0)/T$ from Fig.~1, gives 
$\beta \! = \! 5.36(4)$~mJ/mol~K$^2$T$^{n}$ 
and $n \! = \! 0.75(1)$ at $T \! = \! 2.3$~K,
and $\beta \! = \! 4.70(2)$~mJ/mol~K$^2$T$^{n}$
and $n \! = \! 0.826(8)$ at $T \! = \! 4.2$~K.
Using these two values of the exponent $n$, a linear extrapolation
to $T \! = \! 0$~K gives $n \! = \! 0.66$---which 
is close to the result predicted in Ref.~\cite{Ichioka:99}.
Thus within this theoretical picture, vortex-vortex interactions 
affect the specific heat
by decreasing both the size of the vortex cores and the corresponding density
of bound QP states.
 
A direct test of this idea can be made using the precise field dependence of
the vortex-core radius determined from $\mu$SR. We first note that
for a fixed value of the temperature,
the radius of a coherence length $\xi$ in Eq.~(\ref{eq:Fetter}) 
is usually assumed to be independent of $H$, so that $\xi \approx \xi_0$---in which case the relation 
$\Phi_0 \! = \! 2 \pi \xi^2 H_{c2}$ can be
used to give the familiar result: $C_{\rm cores} \! \sim \! \gamma_n T B/H_{c2}$. 
More generally, however, the density of core states $N(E)$ 
in Eq.~(\ref{eq:DOS}) will have
two sources of magnetic field dependence. First, the vortex density $B(H)/\Phi_0$ 
increases rapidly just above $H_{c1}$ approaching $H/\Phi_0$ as 
$H \! \rightarrow \! H_{c2}$.
Second, due to vortex-vortex interactions, $r_0 \! \approx \! \xi$ 
decreases rapidly just above 
$H_{c1}$ approaching a constant value near $H_{c2}$.
In Ref.~\cite{Sonier:97a}, the fitted value of the GL coherence
length for NbSe$_2$ is well approximated by the 
relation $\xi (H) \! = \! 46(2) \! + \! 28.9(9)/(H \! - \! H_{c1})^{1/2}$   
and $\xi (H) \! = \! 47(5) \! + \! 46(3)/(H \! - \! H_{c1})^{1/2}$ at $T \! = \! 2.3$~K
and $4.2$~K, respectively. Substituting these relations into
Eq.~(\ref{eq:heat}) gives an equation of the form
\begin{equation}
C(T,H) = C(T,0) + c_1 TB(H)
+ \frac{c_2 T B(H)}{(H-H_{c1})^{1/2}} \, ,
\label{eq:form}
\end{equation}
where $c_1$ and $c_2$ are numerical constants.
The form of Eq.~(\ref{eq:form}) is similar to Maki's macroscopic
specific heat equation, since $d^2H_{c1}/dT^2 \! < \! 0$ in Eq.~(\ref{eq:Maki}).
The main difference is that the second term in Eq.~(\ref{eq:form}) was
derived from data at fields $4H_{c1} \! < \! H \! < \! 0.3H_{c2}$, whereas
Eq.~(\ref{eq:Maki}) is valid near $H_{c1}$ only.
 
The form of $B(H)$ in NbSe$_2$ was determined from the 
ZFC magnetization measurements with increasing magnetic field,
where $B(H) \! = \! H \! + \! 4 \pi M$.
A phenomenological equation was chosen to fit the magnetization
data ({\it e.g.}, solid curve in Fig.~2, inset) since there is
no analytical equation from theory which is valid over the 
entire field range. We find that
\begin{eqnarray} 
B (H) \approx & H & - 0.00009(2)H^{-1.32(7)} \nonumber \\
& - & 0.0062(2) \ln (0.91(6)/H) \, , \nonumber \\
B (H) \approx & H &  - 0.00063(2)H^{-0.65(7)} \nonumber \\
& - & 0.00105(12) \ln (0.87(8)/H) \nonumber 
\end{eqnarray}
at $T \! = \! 2.3$~K and $4.2$~K, respectively. 
The contribution of $B(H)$ to the field dependence of $C(H)$ is observed by 
assuming $\xi$ is independent of field and equal to $\xi_0$,
and substituting the relation for $B(H)$ into 
Eq.~(\ref{eq:heat}). The solid curve in the inset of Figs.~3 and 4, is
the corresponding fit which gives $C(T,0)/T \! = \! 9.2(1)$~mJ/mol~K$^2$
and $30.25(5)$~mJ/mol~K$^2$ at $T \! = \! 2.3$~K and $4.2$~K,
respectively. It is clear that    
the field dependence of the magnetic induction
above $H_{c1}$ does not alone account for the 
low-field curvature in $C(H)$. We note that the fitted value of
$C(T,0)/T$ at $T \! = \! 2.3$~K, is nearly $20$~\% larger than that 
measured in zero field.  

To account for the effect of vortex-vortex interactions, we substitute
the precise expression for $\xi(T,H)$ measured by 
$\mu$SR \cite{Sonier:97a} into Eq.~(\ref{eq:heat}) to give
\begin{equation}
\frac{C(T,H)}{T} \! = \! \frac{C(T,0)}{T} 
+ a(T) B(T,H) \frac{[1+b_1(T)~H]}{[1+b_2(T)~H]} \, ,
\label{eq:correct}
\end{equation}
where $a(T) \! = \!  \pi^2 \xi_0 \gamma_n \xi(T,0) I/\Phi_0$, 
$b_1(T)$ and $b_2(T)$ are temperature dependent 
coefficients. The values for $b_1(T)$ and $b_2(T)$ were
determined in the work of Ref.~\cite{Sonier:97a} and are given in
Table~\ref{fits}. Note that $\xi(T,0)$ is merely an extrapolation of
$\xi(T,H)$ from the vortex state to $H \! = \! 0$~T, and should
not be confused with the coherence length in the Meissner state.
The unusual form of the $\mu$SR term has the following interpretation:
({\it i}) the numerator is the field dependence of the fitted magnetic
penetration depth $\lambda_{ab}$ whereas, ({\it ii}) the denominator
is the field dependence of the GL parameter  
$\kappa \! = \! \lambda_{ab}/\xi_{ab}$.

In the main panel of Figs.~3 and 4, the data is plotted over the
field range of the $\mu$SR experiment. The solid curve
is a fit to Eq.~(\ref{eq:correct}), with the fitted
values of $a(T)$ and $C(T,0)/T$ given in Table~\ref{fits}. The 
quality of these fits 
lend strong support to the assertion that the downward curvature 
of $C(H)$ at low fields is partially due to the expansion of the vortex cores. 
Using the fitted values of $a(T)$, $\gamma_n \! = \! 15.7(3)$~mJ/mol~K$^2$
from Fig.~1, the GL value $I \! = \! 1.92$ \cite{Caroli:64},
$\xi(2.3,0) \! = \! 157(4)$~\AA~ and
$\xi(4.2,0) \! = \! 252(10)$~\AA~ from Ref.~\cite{Sonier:97a},
we find that $\xi_0 \! = \! 56(4)$~\AA~ and $\xi_0 \! = \! 42(8)$~\AA~
at $T \! = \! 2.3$~K and $4.2$~K, respectively.
 
As discussed earlier, in a pure $d_{x^2-y^2}$-wave superconductor 
the DOS comes predominantly from the QP spectrum outside of 
the vortex cores, since the presence of the gap nodes inhibit the 
formation of localized states within the cores. In this case
the low-field expansion of the vortex cores recently measured
by $\mu$SR in YBCO \cite{Sonier:97b,Sonier:99}, should
have little effect on $C(H)$. However, the vortex-vortex interactions 
which are responsible for the changing core size should play some
role in the behavior of $C(H)$.
For instance, in Ref.~\cite{Ichioka:99}   
it was shown that due to nearest neighbor vortex-vortex interactions, 
the density of the
extended QP states in a $d_{x^2-y^2}$-wave superconductor
will be proportional to $H^{0.41}$, rather than $\sim \sqrt{H}$, as predicted 
by Volovik \cite{Volovik:93}. Thus far, measurements of $C(H)$ in the HTSCs have 
not been performed with sufficient accuracy to verify this prediction.

In conclusion, we have shown that the 
field dependence of the electronic
specific heat measured in the $s$-wave superconductor NbSe$_2$ is due to 
both the field dependence of the magnetic induction and
the vortex-core size. We attribute the latter contribution to
vortex-vortex interactions.
Our findings provide a simple explanation for the downward
curvature of $C(H)$ reported in superconductors which do not 
have nodes in the energy gap function. 

We thank A.V.~Balatsky, L.~Taillefer and M.~Chiao for helpful 
and informative discussions. The work at
Los Alamos was performed under the auspices of the US Department of Energy.    
%%%%%%%%%%%%%%%%%%%%%%%%%%%%%%%%%%%%%%%%%%%%%%%%%%%%%%%%%%%%%%%%%%%%%%%%%%% 
\newpage 
\begin{center} 
TABLES 
\end{center} 

\begin{table}
\begin{center}
\begin{tabular}{cccccc} 
$T$ & \multicolumn{2}{c}{$C(T,0)/T$} 
& $a$ & $b_1^{\cite{Sonier:97a}}$ & $b_2^{\cite{Sonier:97a}}$ \\
$[$K$]$ & 
\multicolumn{2}{c}{$\left[ \frac{\rm mJ}{\rm mol~K^2} \right]$} 
& $\left[ \frac{\rm mJ}{\rm mol~K^2 T} \right]$ &
$[$T$^{-1}]$ & $[$T$^{-1}]$ \\
 & & & & & \\
 & {\rm Fig.~1} & {\rm Eq.~(\ref{eq:correct})} & & & \\
\hline
2.3 & 7.7(1)  & 7.78(4)  & 12.7(1) & 0.555 & 2.48 \\
4.2 & 29.6(1) & 29.26(9)  & 15.0(5) & 0.891 & 4.68 \\
\end{tabular}
\caption[Paramters]{Results from fitting the specific heat for NbSe$_2$
to Eq.~(\ref{eq:correct}). Only $C(T,0)$ and $a(T)$ were free to vary.}  
\label{fits} 
\end{center}
\end{table} 

\begin{center} 
FIGURE CAPTIONS 
\end{center} 
 
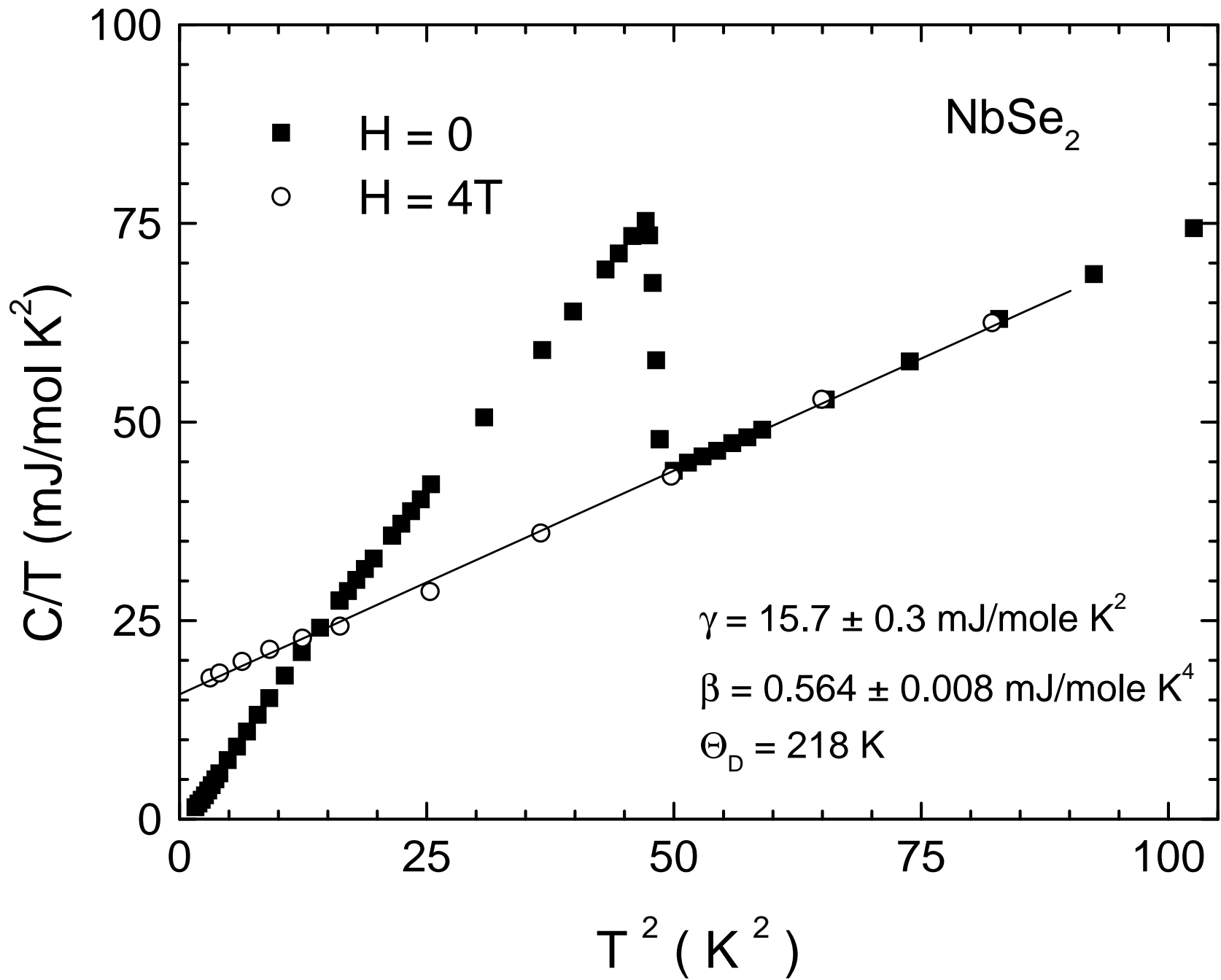
Figure 1. The specific heat of NbSe$_2$ in zero field and
$H \! = \! 4$~T ({\it i.e.} the normal state),
plotted as $C(T,0)/T$ vs $T^2$. The solid line is a fit to
$C_n/T \! = \! \gamma_n \! + \! \beta T^2$, where 
$\Theta_D \! = \! (12 \pi^4 N k_B/5 \beta)^{1/3}$ is the
Debye temperature.\\

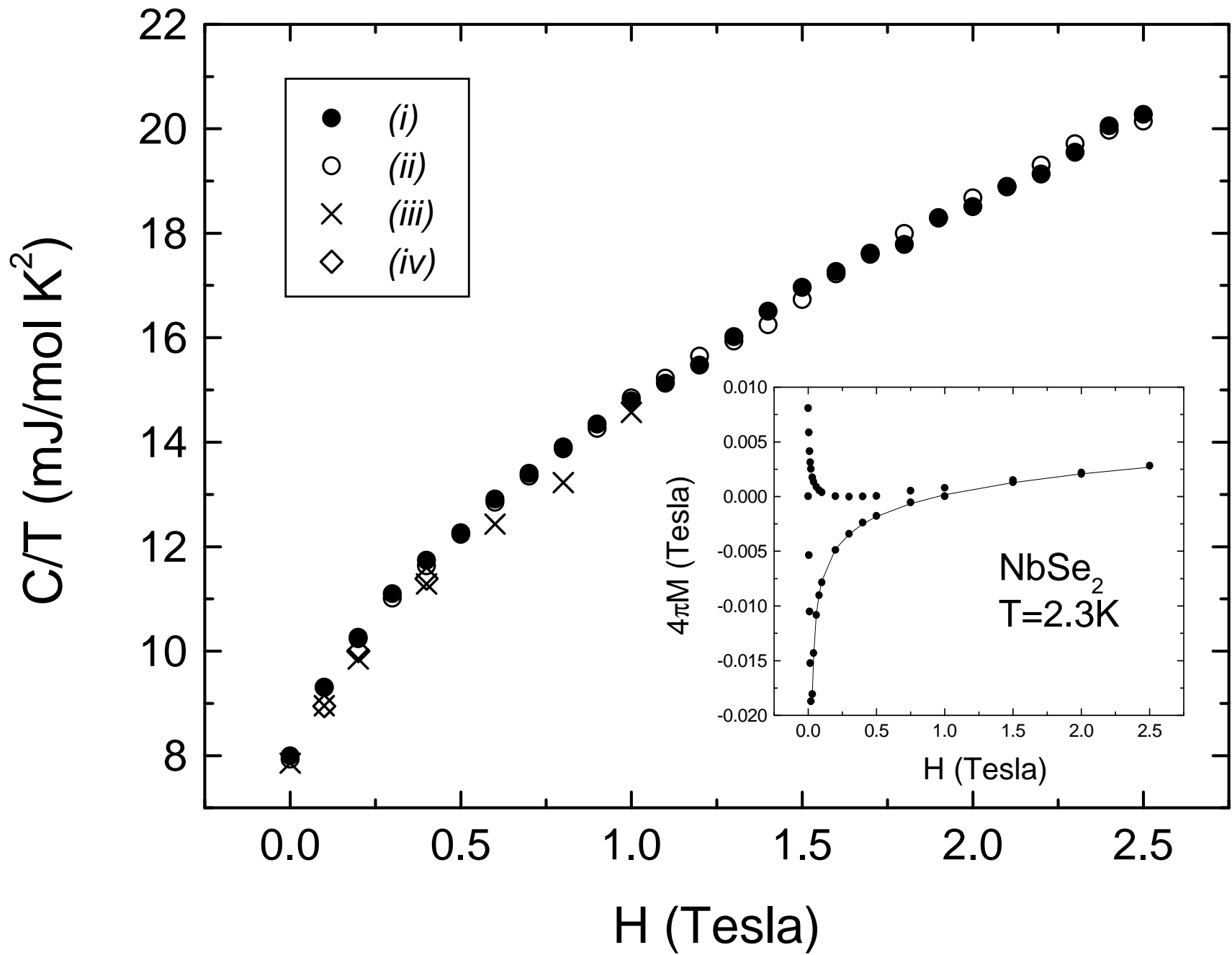
Figure 2. The specific heat of NbSe$_2$ at $T \! = \! 2.3$~K, plotted
as $C(T,H)/T$ vs $H$. The data was taken four different ways, as 
described in the text. The inset shows the DC magnetization-hysteresis 
loop for the ZFC crystal.\\

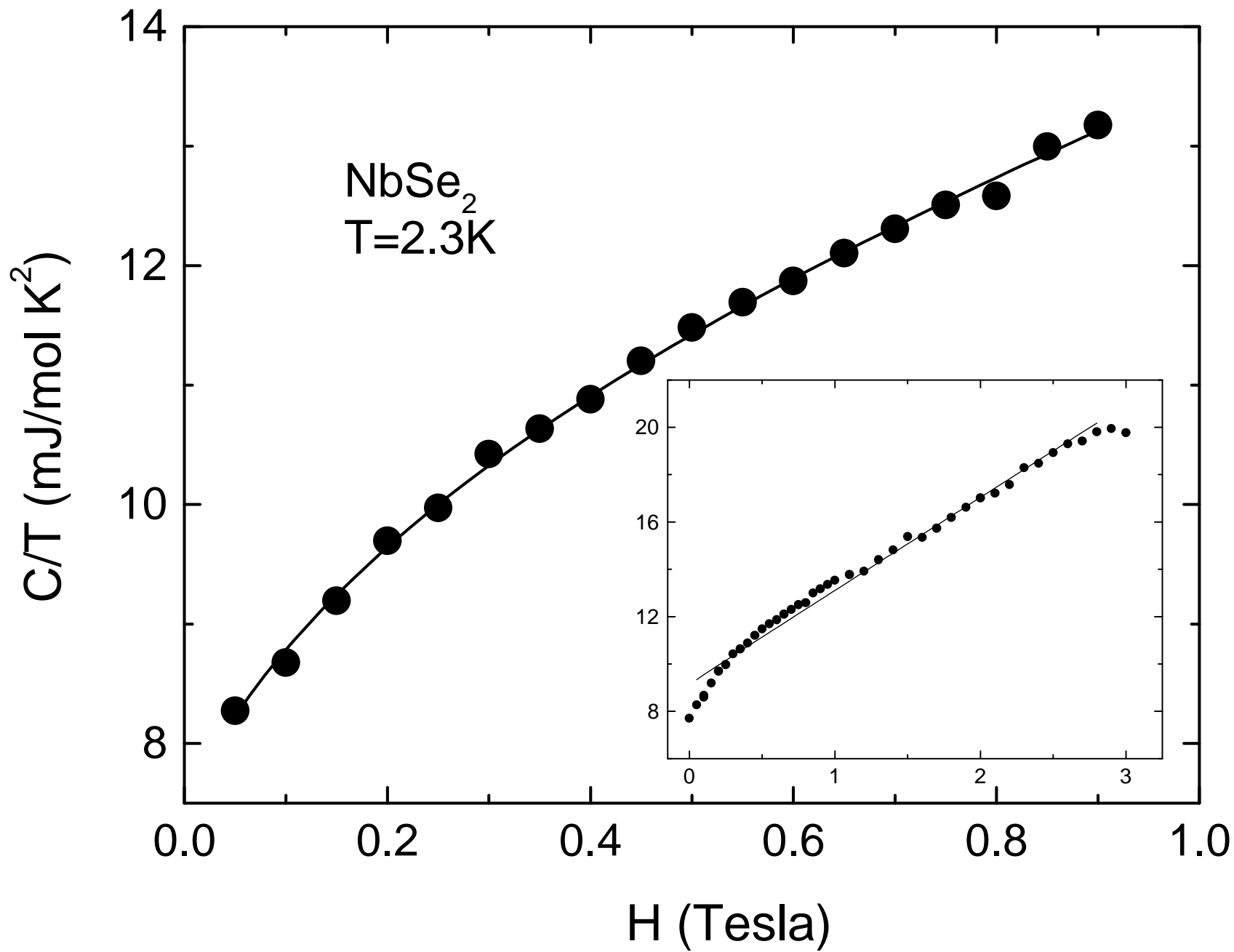
Figure 3. The specific heat of NbSe$_2$ at $T \! = \! 2.3$~K over the
field range of the $\mu$SR experiment \cite{Sonier:97a} plotted 
as $C(T,H)/T$ vs $H$. The solid curve is a fit to Eq.~(\ref{eq:correct}).
Inset: The specific heat over an extended field range. The
solid curve is a fit to Eq.~(\ref{eq:heat}). \\

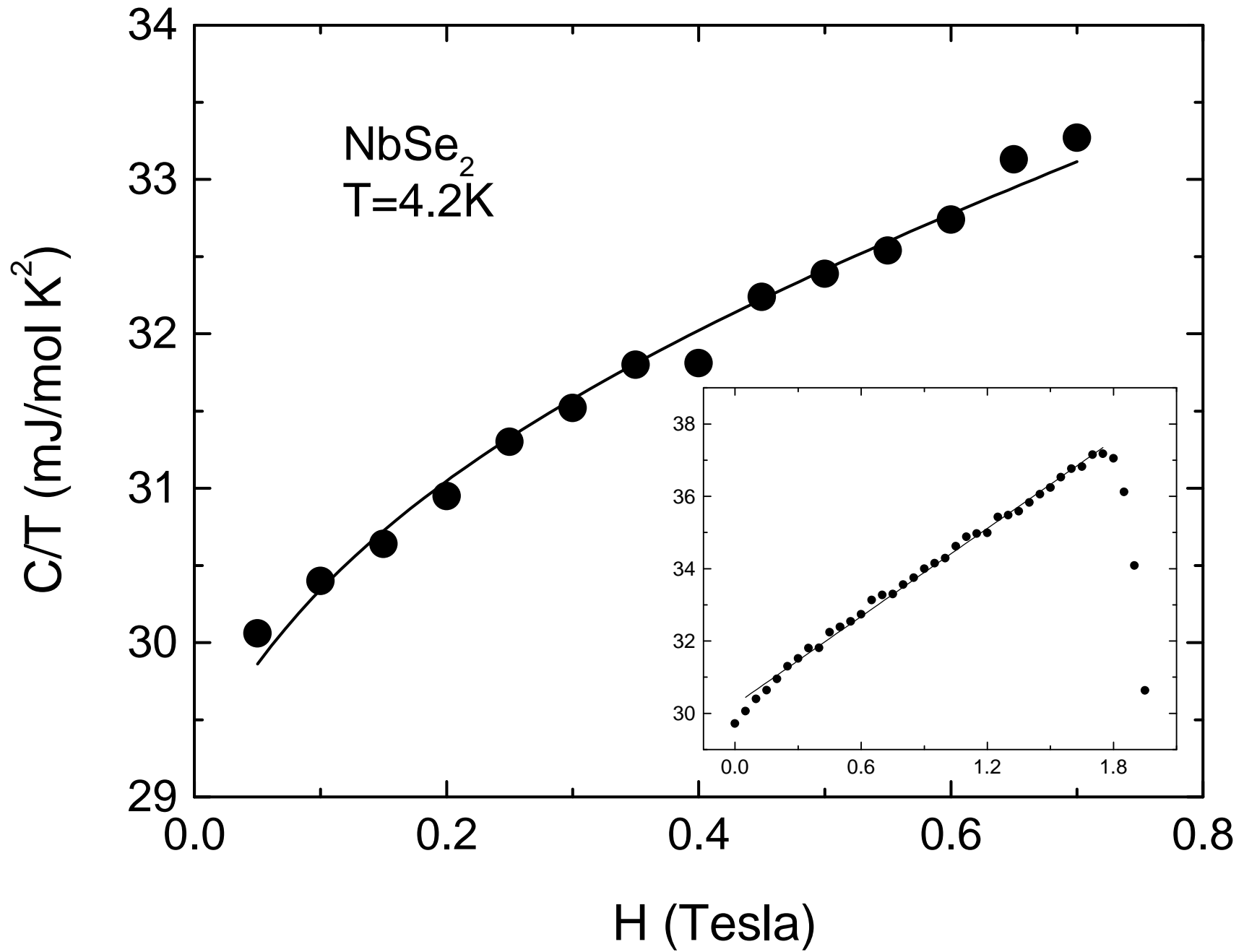
Figure 4. Same as Fig.~3, but with the sample at $T \! = \! 4.2$~K.\\
   
%%%%%%%%%%%%%%%%%%%%%%%%%%%%%%%%%%%%%%%%%%%%%%%%%%%%%%%%%%%%%%%%%%%%%%%%%%%% 
% REFERENCES 
 
\end{document}